\documentstyle[aps,prb,epsfig,graphics]{revtex}

\newcommand{\beq}{\begin{equation}}
\newcommand{\eeq}{\end{equation}}
\newcommand{\beqa}{\begin{eqnarray}}
\newcommand{\eeqa}{\end{eqnarray}}

\begin{document}
\draft

\title{STATISTICS OF CHARGED SOLITONS AND FORMATION OF STRIPES }

\author{S. Teber\protect\( ^{1}\protect \), B. P. Stojkovic\protect\( ^{2}\protect \),
S. A. Brazovskii\protect\( ^{1}\protect \), A. R. Bishop\protect\( ^{2}\protect \)}

\address{\centering \protect\( ^{1}\protect \)Laboratoire de Physique Theorique et
Modeles Statistiques, Bat. 100, Universite Paris-Sud, 91406, Orsay-Cedex, France}

\address{\centering \protect\( ^{2}\protect \)Theoretical Division and Center for Nonlinear
Studies, Los Alamos National Laboratory, Los Alamos, NM 87545}

\maketitle

\begin{abstract}
\noindent The \( M- \)fold degeneracy of the ground state of a quasi-one dimensional
system allows it to support topological excitations such as solitons. We study
the combined effects of Coulomb interactions and confinement due to interchain
coupling on the statistics and thermodynamics of such defects. We concentrate
on a two-dimensional case with \( M=2 \) which may correspond to monolayers
of doped polyacetylene-type polymers or other charge density waves and to junctions
in field effect experiments for equivalent materials. Combining analytical and
numerical methods, the theory is developed by a mapping to the \( 2- \)dimensional
Ising model with long range \( 4 \)-spin interactions. We obtain the phase
diagram depending on the ratio of Coulomb interaction and interchain coupling
with respect to temperature. The latter exhibits deconfined phases for liquids
and Wigner crystals of kinks, and confined ones for bikinks. Also we find aggregated
phases with either infinite domain walls of kinks or finite rods of bikinks.
Roughening effects due to both temperature and Coulomb repulsion are observed.
Applications may concern the melting of stripes in doped correlated materials.
\end{abstract}

\section{Introduction}

\noindent Charged interfaces, from their microscopic origin to their macroscopic
statistical properties, appear naturally in the study of doped correlated insulators.
The latter are characterized by discrete symmetry broken phases with \( M \)-fold
degenerate ground states, \( M>1 \). The interfaces, which are topological
configurations whether charged or not, provide a connection between the different
ground states. In this framework, quasi-one-dimensional systems, e.g. systems
of weakly bound chains, exhibit them at a microscopic level via solitons which
emerge as single particles, the \( \varphi - \)particles \cite{phi,sbhistory}.
This name is related to the fact that each particle provides a phase jump \( \varphi =2\pi /M \),
connecting two different but equivalent ground states, within a chain. Their
presence therefore breaks the coherence between neighboring chains which must
have the same phase modulo \( 2\pi  \). Actually they may exist in a broader
range of materials even though their individual detection might be difficult.
In this respect, \( \pi  \) defects, i.e. \( M=2 \), may be interpreted as
the building blocks of the experimentally observed stripes in manganite \cite{chen}
and cuprate \cite{emery,zaanen} oxides as well as of the domain walls
observed in uniaxial ferroelectric systems \cite{monceau}. The latter references,
\cite{zaanen} and \cite{monceau}, should also provide informations about the
charge carried by kinks via measurements of the tilt angle of the stripes and
the dipole momentum, respectively. 

\noindent In one dimension (\( D \)) or above the transition temperature, uncharged
\( \varphi - \)particles are allowed to exist as a gas of elementary excitations.
They can be defined as strings of reversed spins which, in \( D>1 \), cost
an energy proportional to their length because of interchain coupling. Below
the transition temperature \( T_{c} \), in the ordered broken symmetry phase,
a single soliton would cost infinite energy. The condition for their preservation
lies in the concept of \textit{confinement} \cite{conf}, which limits the string's
length leading to finite energy excitations. \( T_{c} \) is thus called the
confinement - deconfinement transition temperature. Below \( T_{c} \), a constant
force is originated by confinement which binds the kinks into trains of \( M \)
particles, to recover the total increment of \( 2\pi  \), within each chain.
Going to zero temperature, confinement leads to the aggregation of the trains
of solitons belonging to different chains. Thus, the microscopic defects become
mesoscopic complexes which span the whole sample at \( T=0 \) and can be viewed
as domain lines or walls whether the system is, respectively, \( 2D \) or \( 3D \)
\cite{pokr}. 

\noindent These qualitative arguments strictly apply when solitons are uncharged
and must be extended to the case where long ranged Coulomb interactions are
present. Indeed, doping correlated insulators leads to the formation of charged
topological defects \cite{kirova}. Also, field effect experiments \cite{batlogg}
provide junctions of mobile charged kinks interacting with their images. 

\noindent The present work deals exclusively with a \( 2 \)-fold degenerate
ground state which corresponds, for quasi-one-dimensional systems, to the well
known case of polyacetylene, cf. \cite{kirova,su,maki}. In
such (\( M=2 \)) systems the elementary defects are the \( \pm \pi  \)-solitons.
The kink (\( + \)) shifts the phase of the CDW from \( 0 \) to \( \pi  \)
while the anti-kink (\( - \)) shifts it from \( \pi  \) to \( 2\pi  \). Both
amplitude-solitons have the same electric charge. The anti-kink of opposite
charge, corresponding to the \( \pi  \) to \( 0 \) shift, is not thermally
activated. For the sake of clarity we thus attribute to the \( -\pi  \)-soliton
the same graphical representation as the former in Fig. \ref{figsol}. The trains
of kinks are then simply pairs of kink\( - \)anti-kinks or ``bikinks'' and
the domain lines perpendicular to the chains are formed by such bikinks. We
shall examine the statistical properties of kinks under the combined action
of the attractive confinement force and the repulsive \( 3D \) Coulomb force
with the help of both analytical and numerical methods. 

\noindent The paper is organized as follows. In chapter \ref{prent} we present
the general model of confinement based on the mapping of a system of interacting
kinks to the anisotropic Ising model. In chapter \ref{FirstEst} the phase diagram
of the \( 2D \) system, with competing Coulomb interactions and interchain
coupling, is given. Analytically, the results are obtained by various estimates
in the case of strong correlations and with the help of a phenomenological model,
equivalent to the Ising model below \( T_{c} \), which has been extended to
include weak Coulomb interactions. The thermal roughening of uncharged domain
lines, as well as the \( T=0 \) Coulomb roughening, are also considered. In
chapter \ref{numerics}, numerical simulations are presented. Various regimes
of the estimated phase diagram are discussed. The impact of discreteness imposed
by numerical simulations is also considered. Details of calculations are given
in the Appendix.

\section{The general model of confinement}

\label{prent}We shall follow the method of \cite{bohr}, where the statistics
of uncharged kinks in a system of weakly bound chains was studied by \textit{mapping
to the Ising model}. 

\bigskip{}
\noindent The original Hamiltonian describing the kinks reads

\medskip{}
\begin{equation}
\label{h0}
\widetilde{H_{0}}=H_{0}-\mu N_{s}=\int ^{L}_{0}dx\{-\sum _{<\alpha ,\beta >}V\eta _{\alpha }(x)\eta _{\beta }(x)+\sum _{\alpha }(E_{s}-\mu _{s})c_{\alpha }(x)\},
\end{equation}

\medskip{}

\noindent where \( N_{s} \) is the total number of kinks, \( E_{s} \) is the
energy of an isolated kink, \( \mu _{s} \) and \( c_{\alpha }(x) \) are, respectively,
their chemical potential and local concentration. \( V \) is the interchain
coupling constant, chosen to be positive. \( \eta _{\alpha }(x) \) is the order
parameter describing a kink along chain \( \alpha  \). 

\noindent On the atomic scale \( a_{0x} \), the kink extends over a core length
\( \xi _{0}>a_{0x} \), e.g. \( \xi _{0}\approx 7a_{0x} \) for polyacetylene,
around position \( x \). For this system, the microscopic, domain-wall like,
excitation takes the values \( \eta =\pm 1 \) between the kinks and far from
their cores. 

\begin{figure}[tbph]
\begin{center}
\includegraphics[width=3.0in]{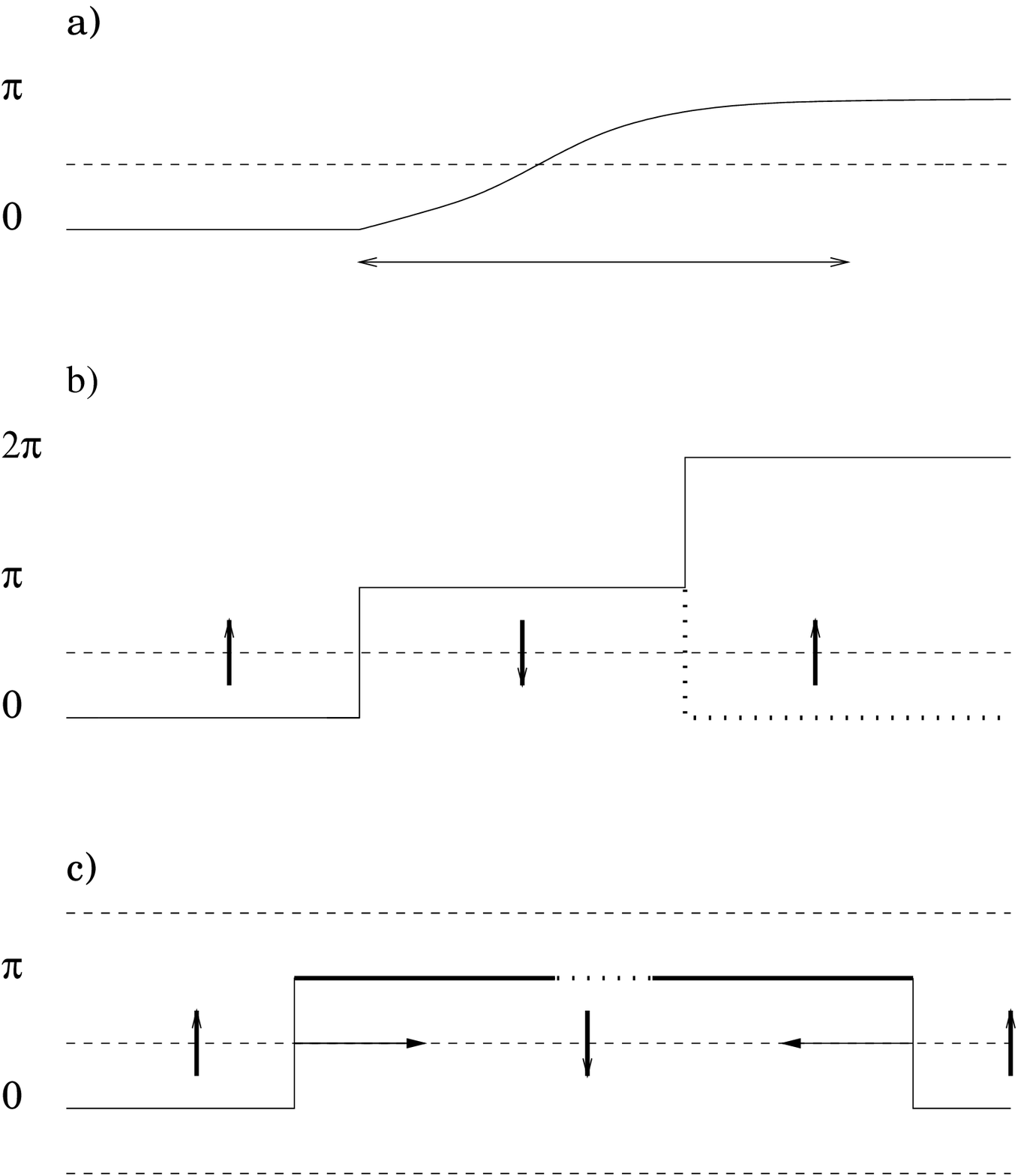}

\caption{\label{figsol}Figure a) displays a \protect\( \pi -\protect \)soliton (solid
line) along a chain (broken line). The arrow indicates the width \protect\( \xi _{0}\protect \)
of the defect. }

\noindent Figure b) displays the soliton and the antisoliton which correspond
to strings of inverted Ising spins after coarse graining. The anti-soliton corresponding
to a shift of the phase from \( \pi  \) to \( 2\pi  \) has the same charge
as the soliton. The other anti-soliton (dotted line) has an opposite charge
and is not thermally activated. 

\noindent Figure c) displays a soliton and an antisoliton of the same charge.
For simplicity the \( \pi  \) to \( 2\pi  \) kink has been represented as
the \( \pi  \) to \( 0 \) one which does not exist in our case. Neighboring
chains (broken lines) are supposed to be free from defects. Due to interchain
interactions (bold lines) the energy of the bikink increases linearly with its
size. As a consequence, a constant confinement force binds the solitons. Exponential
core repulsion prevents them from annihilating.
\end{center} 
\end{figure}

\noindent The mapping to the two-dimensional Ising model, whose \( Z_{2} \)
symmetry is clearly compatible with the \( 2- \)fold degeneracy of our system,
is then based on a coarse graining procedure by introducing a new length scale
\( a_{x}>\xi _{0}>a_{0x} \), which corresponds to the optimal width of a pair
of kinks within a chain under the combined action of interchain interaction
and exponential intrachain repulsion between them, cf. Fig. \ref{figsol}. We
then discretize the system by \( x\rightarrow x_{n}=na_{x} \) and \( y\rightarrow y_{\alpha }=\alpha a_{y} \)
where \( a_{y} \) is the interchain distance. On this scale the order parameter
can be restricted to the values \( \pm 1 \) so it is written as an Ising spin,
\( S_{n,\alpha } \). The kink density, corresponding to the number of kinks
per site, can also be expressed in terms of the Ising spins in the following
way. The kink, on chain \( \alpha  \) for example, is defined \textit{between}
two consecutive opposite spins, say on sites \( n \) and \( n+1 \), cf. Fig.
\ref{figsol}. The dimensionless density of solitons \( \rho _{n,\alpha } \)
is related to their dimensional concentration \( c_{\alpha }(x) \), cf. (\ref{h0}),
by

\medskip{}
\begin{equation}
\label{rho}
c_{\alpha }(x)a_{x}=\rho _{n,\alpha }.
\end{equation}

\medskip{}

\noindent In terms of spins these considerations yield

\medskip{}
\begin{equation}
\label{rho/spin}
\rho _{n,\alpha }=\frac{1}{2}(1-S_{n,\alpha }S_{n+1,\alpha }).
\end{equation}

\medskip{}

\noindent We see that \( \rho _{n,\alpha } \) is unity when a kink is present
and zero otherwise.

\bigskip{}
\noindent Introducing these new variables in the first Hamiltonian (\ref{h0})
and including Coulomb interactions we recover the two-dimensional Ising model 

\medskip{}
\begin{equation}
\label{Htot}
\widetilde{H}=\widetilde{H_{0}}+H_{c}=-J_{\perp }\sum _{<\alpha ,\beta >n}S_{n,\alpha }S_{n,\beta }-J_{\parallel }\sum _{\alpha ,n}(S_{n,\alpha }S_{n+1,\alpha }-1)+H_{Coul},
\end{equation}

\smallskip{}
\noindent where 

\smallskip{}
\begin{equation}
\label{IsingParam}
J_{\perp }=Va_{x}=K_{\perp }T\qquad \qquad J_{\parallel }=\frac{1}{2}(E_{s}-\mu _{s})=K_{\parallel }T.
\end{equation}

\smallskip{}

\noindent The first term in (\ref{Htot}) describes the interchain interaction,
responsible for the confinement of the kinks, whose strength is given by the
dimensionless coupling constant \( K_{\perp } \). In the second term, \( K_{\parallel } \)
is the effective chemical potential of a kink, which has been included as a
coupling constant. We emphasize the fact that the energy of a free soliton \( E_{s} \)
is always larger than the temperature, so that none of them is thermally activated.
We only consider redistributions of the kinks, no matter how they have been
created (doping, charge transfer, incommensurabilities), under the combined
action of confinement and Coulomb interactions. The latter has been introduced
in (\ref{Htot}) through

\medskip{}
\begin{equation}
\label{Hc}
H_{Coul}=\frac{e^{2}}{2\epsilon }\sum _{n,m;\alpha ,\beta }\frac{(\rho _{n,\alpha }-\nu )(\rho _{m,\beta }-\nu )}{|\overrightarrow{r}_{n,\alpha }-\overrightarrow{r}_{m,\beta }|},
\end{equation}

\medskip{}

\noindent where \( \overrightarrow{r}_{n,\alpha }=(na_{x},\alpha a_{y}) \),
\( \nu =\overline{c_{\alpha }(x)}a_{x} \) is the average dimensionless density
of kinks, see (\ref{mu}) below, and \( \epsilon  \) the dielectric constant
of the isotropic, neutral media in which the plane is embedded. The Coulomb
interaction as written above takes into account the presence of a homogeneous,
negatively charged, background of density \( \nu  \) which ensures the electroneutrality
of the whole system. It gives the total electrostatic energy of the kinks and
the background they are interacting with. The case of dipoles, produced in field
effect junctions will be considered later.

\bigskip{}
\noindent As we are working in the grand-canonical ensemble, the key for finding
the thermodynamic properties of the system, is to determine the chemical potential
of the kinks via the average density per site of kinks:

\medskip{}
\begin{equation}
\label{mu}
\nu =\frac{1}{2}(1-<S_{n,\alpha }S_{n+1,\alpha }>)\equiv \frac{1}{2}\left( 1+\frac{1}{T}\frac{\partial f_{I}}{\partial K_{\parallel }}\right) \equiv -\frac{\partial \Omega _{s}}{\partial \mu _{s}},
\end{equation}
where the averaging \( <.> \) and the free energy \( f_{I} \) are defined
for the ensemble of Ising spins and \( \Omega _{s} \) is the grand potential
of the solitons, cf. \cite{bohr}. 
\medskip{}

\section{The two-dimensional system}

\label{FirstEst}

\subsection{\noindent The dimensionless parameters of the phase diagram}

\noindent Without Coulomb interaction, the Ising model mapping allows for a
complete analysis of the \( 2D \) case. The aggregation of kinks is controlled
by the temperature or the reduced temperature used in the phase diagram,
\[
\Theta =\frac{T}{J_{\perp }}.\]
If their density is weak enough, which we will suppose in the following, below
the confinement transition temperature
\begin{equation}
\label{tetac}
T_{c}\approx \frac{J_{\perp }}{\nu }\qquad ,\qquad \Theta _{c}\approx \frac{1}{\nu }
\end{equation}
the growing confinement energy binds the kinks in pairs with remaining energy
\( J_{\perp } \). When the bikinks on neighboring chains stick together below
the crossover temperature,
\begin{equation}
\label{teta0}
T_{0}\approx \frac{4J_{\perp }}{\log (2\nu )}\qquad ,\qquad \Theta _{0}\approx \frac{4}{\log (2\nu )},
\end{equation}
the energy \( J_{\perp } \) is released which results in an adhesion process
leading to growing rods. The latter are double lines formed by aggregated bikinks,
perpendicular to the chains. Their growth when temperature is further decreased
can be seen from the density of solitons and the average rod length \( l_{\perp } \)
\begin{equation}
\label{v02}
\nu =\frac{\exp (-4K_{\perp })}{8K^{2}_{\parallel }}\qquad \qquad l_{\perp }=\frac{a_{y}}{2K_{\parallel }}=a_{y}\sqrt{2\nu }\exp (2K_{\perp }).
\end{equation}
These expressions, obtained from the exact solution of the non-interacting Ising
model \cite{bohr} show that as \( T\rightarrow 0 \) the length of the aggregates
grows exponentially with temperature until they cross the whole sample more
like a system of free domain lines. 

\bigskip{}
\noindent On the other hand, the strength of the Coulomb interaction can be
characterized by 

\medskip{}
\begin{equation}
\label{gamma}
\Gamma =\frac{w_{y}}{J_{\perp }}\qquad \qquad w_{y}=\frac{(ze)^{2}}{\epsilon a_{y}}
\end{equation}

\medskip{}

\noindent where \( s=a_{x}a_{y} \) is the unit area of the coarse grained lattice
and \( z<1 \) takes into account possible fractional charges \cite{kirova}
which may be measured in ferroelectric experiments \cite{monceau}. Parameters
\( J_{\perp } \)and \( w_{y} \) give, respectively, the scale of the confinement
energy and of the Coulomb energy perpendicular to the chains. They are invariant
under coarse graining, thus \( \Gamma  \) is also invariant. Unless \( \epsilon  \)
is very large the \( w_{y} \) scale is larger than other ones. For example,
it is much larger than the energy per particle of a Wigner crystal, which is
of the order of \( e^{2}/\epsilon R \) where \( R\gg a_{x,y} \) is the mean
distance between particles of the Wigner crystal. It is also important with
respect to the confinement. In Rydbergs, \( 1Ry=e^{2}/\epsilon a_{0} \), where
\( a_{0} \) is Bohr's radius, we have \( w_{y}=za_{0}Ry/\epsilon a_{y} \).
In the same units, \( J_{\perp }\approx \nu T_{c}\approx \nu 10^{-3}Ry \) if
we choose \( T_{c}\approx 100K \). Thus for \( a_{y}=5a_{0} \) and \( \nu =1\% \)
we have
\begin{equation}
\label{GammaNumerical}
\Gamma =\frac{w_{y}}{J_{\perp }}\approx z10^{2}/\epsilon .
\end{equation}
Taking \( z=1 \), the dielectric constant needs to be of the order of \( 10^{2} \)
for the confinement energy to be of the order of the Coulomb energy. This is
large but may not be excluded for \( \epsilon \sim (\omega _{p}/\Delta )^{2} \),
where the plasma frequency \( \omega _{p}\approx 10^{-1}-10 \) eV and the gap
\( \Delta \approx 10^{-2}-10^{0} \) eV.

\noindent At low temperatures \( \Gamma  \) controls the aggregation of the
defects in a way similar to \( \Theta  \) at high temperatures. This follows
from the derivation of the optimal width \( x_{opt} \) of a bikink under the
combined action of Coulomb force \( e^{2}/\epsilon x^{2} \) and the confinement
force \( -V \):
\begin{equation}
\label{xoptbikink}
x_{opt}=\sqrt{s\Gamma }
\end{equation}
This relation holds in all bikink phases whether they are liquid or of the Wigner
crystal type. In both phases the average distance between bikinks is given by
\( R=\sqrt{s/\nu } \). As long as \( x_{opt}\ll R \) the elongation \( x_{opt} \)
can be neglected so that, in zero approximation, the bikink can be considered
as a well defined particle of charge \( 2e \). On the other hand when \( x_{opt} \)
is of the order of \( R \), bikinks deconfine and the system consists of interacting
solitons. These remarks are weakly modified at non zero temperature as long
as \( \Theta \ll 1 \). This follows from the fact that thermal fluctuations
shift the mean width of the bikink to \( \overline{x}\approx x_{opt}+a_{x}\Gamma ^{1/4}\Theta ^{1/2} \).
Thermal fluctuations can be neglected as soon as \( \Gamma \gg \Theta ^{2} \).
Unless \( \epsilon  \) is large, cf. (\ref{GammaNumerical}), this condition
is well satisfied below \( \Theta _{0} \).

\subsection{Wigner crystals of kinks and bikinks }

Below \( \Theta _{0} \), when the condition \( \Gamma \gg \Theta ^{2} \) is
satisfied, bikinks order in the form of a Wigner crystal. The ratio \( x_{tot}/R=\sqrt{\Gamma \nu } \)
is another control parameter which defines a crossover to a deconfined phase
consisting of a Wigner crystal of solitons. The Lindeman criterion specifies
that the crossover occurs when the mean square deviation of the width \( x_{opt} \)
becomes of the order of a fraction of \( R \). With the small Lindeman number
\( \gamma  \) the transition is thus expected at
\begin{equation}
\label{gc}
\Gamma _{c}\sim \gamma /\nu ,
\end{equation}

\noindent which scales as \( \Theta _{c} \). 

\noindent This deconfinement transition with respect to the bikinks can be easily
derived from the Wigner crystal phase of solitons. The energy per particle of
the Wigner crystal of solitons is \( -e^{2}/\epsilon R \) and the confinement
energy \( \sim J_{\perp }R/a_{y} \). A confinement crossover, with respect
to the solitons, takes place when both energies are of the same order. Taking
into account the Lindeman number, we recover (\ref{gc}).

\noindent These results are summarized in the phase diagram of figure \ref{Fig2.phd}.

\subsection{\noindent The interacting rod phase }

\label{weakCoul}The Coulomb interactions are expected to be especially important
for aggregated phases, e.g. rods and lines, because of the size-dependent charging
energy cost of the aggregation, 
\begin{equation}
\label{sigma}
\Sigma (n)\approx 2w_{y}n\log (\frac{min(na_{y},l_{D})}{a_{x}}),
\end{equation}
where \( n \) is the number of solitons belonging to each of the two lines
formed by the rods and \( l_{D} \) is the Debye screening length. The latter
is determined by the density response function \( \partial \nu /\partial \mu _{s} \)
which, for the \( 2D \) plane embedded in \( 3D \) space, is
\begin{equation}
\label{ld}
\frac{1}{l_{D}}=\frac{2\pi e^{2}}{\epsilon s}\frac{\partial \nu }{\partial \mu _{s}}.
\end{equation}

\noindent We shall consider the case of relatively weak Coulomb interactions
which does not prevent the actual existence of the aggregates while affecting
their statistical properties. The required condition is that \( \Gamma \ll 1 \).
Our aim is then to formulate the general expressions which yield the density
of kinks and their mean length. This can be done with the help of a phenomenological
model which has been shown to adequately reproduce the results of the exact
solution \cite{bohr}. The rods can fuse into longer ones or disintegrate into
shorter ones, equilibrating their chemical potentials. The main effect of the
Coulomb interaction is to shift the energy \( E(n) \) of each aggregate of
length \( na_{y} \) by its self-energy (\ref{sigma}) while preserving their
shapes. The free energy of the system is then
\begin{equation}
\label{frod}
f_{rod}=\sum ^{\infty }_{n=1}\left( c(n)E(n)+Tc(n)\log (c(n)/e)\right) ,
\end{equation}

\noindent where \( c(n) \) is the density of rods of length \( na_{y} \),
the second term is the entropy of the Boltzmann-like rod gas, and 
\begin{equation}
\label{energy}
E(n)=4J_{\perp }+2E_{s}n+\Sigma (n)
\end{equation}

\noindent is the energy of a rod of length \( na_{y} \). The first term in
(\ref{energy}) corresponds to the confinement energy paid by the tips of the
rod. In the second term we have taken into account the fact that there are two
kinks per unit length of the aggregate. 

\noindent Minimizing the free-energy with the constraint that the total number
of particles is constant,
\begin{equation}
\label{nu/derivfrod}
0=\frac{\partial }{\partial c(n)}\left( f_{rod}-\mu _{s}\sum _{l}2nc(n)\right) ,
\end{equation}

\noindent the density of kinks is then given by 
\begin{equation}
\label{mu/n}
\nu =\sum ^{\infty }_{n=1}2nc(n)=e^{-4K_{\perp }}\sum ^{\infty }_{n=1}2n\exp \left( -4nK_{\parallel }-\beta \Sigma (n)\right) ,
\end{equation}
where \( K_{\parallel } \) is related to the chemical potential of the solitons
\( \mu _{s} \) by (\ref{IsingParam}). Thus, expression (\ref{mu/n}) allows
one to fix \( \mu _{s} \). Finally, the average length of the domain line reads
\begin{equation}
\label{l*}
l_{\perp }=a_{y}\frac{\nu }{c}\qquad \qquad c=\sum _{n=1}^{\infty }c(n).
\end{equation}
From (\ref{mu/n}) and (\ref{l*}) with \( \Sigma (n)\equiv 0 \) we recover
the results in (\ref{v02}). As will be shown in subsequent paragraphs, screening
by normal carriers, cf. \( l_{D} \), is important in the present regime. Self-screening
in the ensemble of solitons is also considered. It plays a dominant role at
\( T=0 \), especially for stronger Coulomb interactions corresponding to \( \Gamma \gg 1 \).

\subsubsection{The case of efficient screening}

\noindent The efficiency of the bulk screening is manifested by the fact that
the Debye length \( l_{D} \) is shorter than the average length of the domain
line \( l_{\perp } \). Considering a distribution of rods with lengths \( na_{y}>l_{D} \),
we arrive, with the help of (\ref{sigma}) and (\ref{mu/n}), at the expressions
given in (\ref{v02}) with \( K_{\parallel }\rightarrow K^{*}_{\parallel } \),
where
\begin{equation}
\label{Kpara*}
K^{*}_{\parallel }=K_{\parallel }+\frac{1}{2T}w_{y}\log (\frac{l_{D}}{a_{x}})
\end{equation}
is the effective chemical potential shifted by the Coulomb self-energy. We can
now check our original hypothesis \( l_{\perp }>l_{D} \). We have with the
help of (\ref{ld})
\begin{equation}
\label{ldspecial}
l^{-1}_{D}=\frac{2\pi e^{2}}{\epsilon s}\frac{\exp (-4K_{\perp })}{T}\frac{1}{(2K^{*}_{\parallel })^{3}}.
\end{equation}
Thus

\noindent 
\[
\frac{l_{\perp }}{l_{D}}\approx \frac{2\pi e^{2}}{a_{x}}\frac{\exp (4K_{\perp })}{T}(2\nu )^{2},\]

\noindent which grows as \( T\rightarrow 0 \), i.e. \( l_{\perp }/l_{D}\rightarrow \infty  \).
Therefore, thanks to the presence of an effective bulk screening, the interacting
system can be mapped to the non-interacting one, in the limit \( K_{\perp }\gg 1 \),
proving the growing macroscopic aggregation of the defects.

\subsubsection{The case of weak screening}

More surprising is the case of \textit{weak} screening which corresponds to
\( l_{\perp }<l_{D} \). Considering a distribution of rods with lengths \( na_{y}<l_{D} \),
(\ref{mu/n}) then becomes
\begin{equation}
\label{nnonscreen}
\nu \approx e^{-4K_{\perp }}\int ^{l_{D}}_{0}dn2n\exp \left( -4nK_{\parallel }-2n\frac{w_{y}}{T}\log (\frac{na_{y}}{a_{x}})\right) 
\end{equation}

\noindent where the second term in the exponent comes from the self-energy of
a rod of length \( n \). Due to the non-linearity of the self-energy the chemical
potential cannot simply be shifted as in the previous case. Moreover, the Coulomb
energy makes the sum converge even for a zero value of the chemical potential.
In order to balance the Coulomb interaction and thus keep a given non-zero \( \nu  \),
we must allow \( K_{\parallel } \) to acquire negative values in (\ref{nnonscreen}).
To show this, we notice that the integral in (\ref{nnonscreen}) can be evaluated
with the help of a saddle point approximation, which is valid for large, negative
\( K_{\parallel } \). At the saddle point the argument of the exponential in
(\ref{nnonscreen}) has a sharp maximum for

\noindent 
\begin{equation}
\label{lsp}
l_{sp}\approx \exp \left( \frac{2T|K_{\parallel }|}{w_{y}}\right) 
\end{equation}
 and (\ref{nnonscreen}) becomes
\begin{equation}
\label{vsp}
\nu \exp \left( 4K_{\perp }\right) \approx 2\sqrt{\frac{Tl_{sp}}{w_{y}}}\exp \left( \frac{w_{y}}{T}l_{sp}\right) .
\end{equation}
From (\ref{ld}) and (\ref{nnonscreen}) at the saddle point, the Debye length
reads
\begin{equation}
\label{LspLd}
\frac{1}{l_{D}}\approx \frac{w_{y}}{Ta_{x}}\nu l_{sp}.
\end{equation}
As we work at constant \( \nu  \), \( l_{sp} \) is given by (\ref{vsp})

\noindent 
\begin{equation}
\label{lspsps}
l_{sp}\approx 4\frac{J_{\perp }}{w_{y}}+\frac{T}{w_{y}}\log \left( \frac{\nu }{4}\frac{w_{y}}{\sqrt{J_{\perp }T}}\right) .
\end{equation}
Due to the weak Coulomb interaction \( \Gamma \ll 1 \) and the sharpness of
the saddle point ensured by \( w_{y}\gg T \), the second term in (\ref{lspsps})
is a correction with respect to the first. The second term is responsible for
the increase of \( l_{sp} \) with decreasing temperature as long as 
\begin{equation}
\label{Tl}
T>\frac{\nu ^{2}w^{2}_{y}}{16J_{\perp }}=T_{l}.
\end{equation}
Hence, in this regime, \( l_{sp}\approx 4J_{\perp }/w_{y} \) and (\ref{LspLd})
yields
\begin{equation}
\label{ls/ld}
\frac{l_{sp}}{l_{D}}\approx \frac{16J^{2}_{\perp }}{Tw_{y}}\nu .
\end{equation}
When \( l_{sp} \) reaches \( l_{D} \) a crossover from the actual process
of growth, involving negative \( K_{\parallel } \), to the previous one will
take place at \( T_{cr} \) defined, with the help of (\ref{ls/ld}), by
\begin{equation}
\label{Tcr}
\frac{T_{cr}}{T_{c}}\approx 16\nu ^{2}\frac{J_{\perp }}{w_{y}},
\end{equation}
where (\ref{tetac}) has been used. From (\ref{Tl}) and (\ref{Tcr}) we have
\begin{equation}
\label{Tcr/Tl}
\frac{T_{cr}}{T_{l}}\approx \frac{16^{2}}{\nu }\left( \frac{J_{\perp }}{w_{y}}\right) ^{3}\gg 1,
\end{equation}

\noindent where we have also used the fact that we are in the weak Coulomb regime.
Thus (\ref{Tcr/Tl}) indicates that the well screened regime ( \( l_{sp}>l_{D} \))
is reached before temperature is low enough to disfavor the growth of the rods,
i.e. (\ref{Tl}) is satisfied. We summarize the case of weak screening as follows.

\noindent Without Coulomb interactions, the rods started to form at \( T<T_{0}\approx J_{\perp }/|\log \nu | \)
and then grew exponentially with an activation energy \( 4J_{\perp } \); as
temperature approached zero, the chemical potential approached zero exponentially,
cf. (\ref{v02}).

\noindent In the case of weak Coulomb interaction (\( w_{y}<J_{\perp } \)),
at low temperatures (\( T\ll w_{y} \)) the behavior of the system is very different.
Remarkably, negative \( K_{\parallel } \) corresponds to the Ising model with
antiferromagnetic coupling along the chains. Nevertheless, \( 4- \)spin Coulomb
interactions (\ref{Hc}) keep the system in the ferromagnetic state so as to
preserve electroneutrality. The average length of the aggregates \( l_{sp} \)
is approximately \( J_{\perp }/w_{y} \) with weak deviations linear in temperature
up to some logarithmic corrections, cf. (\ref{lspsps}). These deviations lead
to an increase of \( l_{sp} \) when \( T \) decreases. The chemical potential
\( T|K_{\parallel }| \) moves down within the negative regime but logarithmically
slow as can be seen from (\ref{lsp}). Meanwhile the compressibility \( \approx \partial \nu /\partial \mu _{s} \)
increases so that the screening length decreases linearly with temperature,
cf. (\ref{LspLd}). At \( T_{cr} \) the Coulomb interaction becomes screened.
The system then returns to the exponential growth of \( l_{\perp } \) as for
the noninteracting system.

\subsubsection{Application to field effect junctions}

To finish with this section we mention that the logarithmically growing self-energy
is peculiar to a doped system. In the case of a layer produced by a field effect
experiment, dipoles order in lines of length \( a_{y}n \) and the associated
energy is
\[
E_{dip}(n)=\frac{(ed)^{2}}{2a^{2}_{y}}n\left( \frac{1}{a^{2}_{x}}-\frac{1}{(a_{y}n)^{2}}\right) ,\]
where \( d \) is the distance between a charge and its image. As we are interested
in large \( n \), this energy becomes linear in \( n \). Thus this case is
similar to the well screened case.

\subsection{Thermal and Coulomb roughening of domain lines}

\subsubsection{Thermal roughening of lines}

The effect of thermal fluctuations on the aggregates of the non-interacting
system can be derived from the solid-on-solid (SOS) model which is well known
in the field of random interfaces, cf. \cite{forg} for more details. In this
model a line made of one topological charge, i.e. a domain line, is described
by integer-valued height variables \( x_{k} \) (\( k=1,...,l_{\perp } \))
which define the position of kinks along the chains with respect to their \( T=0 \)
position. The energy of an SOS interface at \( T\rightarrow 0 \), i.e. \( K_{\perp }\rightarrow \infty  \)
and \( l_{\perp }\rightarrow \infty  \), can then be taken as \cite{forg}
\begin{equation}
\label{ESSOS}
\beta H_{SOS}=2K_{\perp }\sum ^{l_{\perp }}_{k=1}|x_{k}-x_{k-1}|.
\end{equation}
 The difference correlation function is then \cite{forg}
\begin{equation}
\label{SOSCorrF}
<(x_{k}-x_{j})^{2}>_{SOS}=2|k-j|\frac{\exp (-2K_{\perp })}{1+2\exp (-2K_{\perp })}.
\end{equation}
Expression (\ref{SOSCorrF}) shows that the correlation function diverges in
the limit \( |k-j|\rightarrow \infty  \). This is a signature of the \textit{roughening}
of the domain line at any non zero temperature. This roughening leads to collisions
between different lines separated by a distance \( a_{x}/\nu  \). Taking into
account the fluctuations of one line, neighboring lines remaining straight for
simplicity, a collision takes place once the mean fluctuation in \( x- \)direction
is of the order of \( a_{x}/\nu  \). Collisions of lines can be equivalently
interpreted as disintegration of rods as can be seen from (\ref{SOSCorrF}).
With \( <(x_{k}-x_{j})^{2}>_{SOS}\approx 1/\nu ^{2} \), the temperature at
which rods of length \( l_{\perp }\approx |k-j| \) are formed is given by
\begin{equation}
\label{T}
T\approx \frac{2J_{\perp }}{|\log \left( l_{\perp }\nu ^{2}\right) |}.
\end{equation}
 Equation (\ref{T}) shows that rods reduce to bikinks (\( l_{\perp }\approx 1 \))
when \( T\geq T_{0} \) in agreement with what has been said in the preceding
paragraphs.

\subsubsection{Coulomb roughening of lines}

\noindent In previous sections we have shown that the growing of the aggregates
as we approach zero temperature is mainly due to an effective bulk screening.
However, at \( T=0 \) all kinks have aggregated into domain lines which cross
the whole sample. Thus, no screening charge remains and no confinement energy,
due to interchain interactions, is paid by the lines. The latter implies that
two domain lines, each made of one type of topological charge, will be far apart.
On the other hand, because there is no more screening charge, the competition
between Coulomb interaction and confinement leads to \textit{roughened lines,}
as shown below\textit{.} It is a signature of the \textit{self-screening} of
the domain lines. This roughening may be considered separately from the thermal
roughening we discussed previously.  

\noindent First, we give some simple estimates concerning the form of the line.
The latter corresponds to a linear distribution of charges and it's equilibrium
shape is determined by allowing one of it's constituent kinks to find it's equilibrium
position under the combined action of the Coulomb repulsion and the attractive
confinement force. The Coulomb force \( F_{coul}=w_{y}/x \) originates from
the infinite line and is balanced by the confinement force \( F_{conf}=-V \).
The optimal equilibrium position of a kink emitted from the line is
\begin{equation}
\label{xopt/2DCoul}
x_{0}\approx \frac{w_{y}}{J_{\perp }}a_{x}=\Gamma a_{x}.
\end{equation}

\noindent For \( \Gamma \gg 1 \), the linear distribution acquires a width
\( x_{0}\gg a_{x} \). This finite width is a manifestation of the self-screening
of the rods which leads to their roughening. 

\noindent A particular distribution of kinks \textit{within} the roughened domain
line requires a more elaborate analysis. We consider a general distribution
characterized by the variational parameters \( x_{0} \) and \( y_{0} \). The
latter correspond to the periodicity of the distribution while the former is
the modulation amplitude. They are found by minimizing the total energy per
chain, \( E_{tot} \), of the distribution. The latter is

\begin{equation}
\label{TotalEnergy}
E_{tot}=E_{coul}+E_{conf}
\end{equation}
with the Coulomb energy per chain
\begin{equation}
\label{poten}
E_{coul}=\frac{w_{y}}{L}\int ^{L}_{0}\frac{dydy'}{a^{2}_{y}}\frac{1}{\sqrt{(y-y')^{2}+\left( x(y)-x(y')\right) ^{2}}},
\end{equation}
where \( L \) is the size of the system in the \( y \) direction. The confinement
energy per chain is

\[
E_{conf}=\frac{a_{y}}{y_{0}}\frac{J_{\perp }}{a_{x}}\int ^{y_{0}}_{0}|\frac{dx}{dy}|dy.\]

\noindent In the case \( x_{0}\ll y_{0} \), (\ref{eCoul<<res}) in the Appendix
and (\ref{TotalEnergy}) yield in the main logarithmic approximation
\begin{equation}
\label{EtotInf}
E_{tot}\approx -2w_{y}\left( \frac{x_{0}}{y_{0}}\right) ^{2}\log \left( \frac{y_{0}}{a_{x}}\right) +4a_{y}\frac{J_{\perp }}{a_{x}}\frac{x_{0}}{y_{0}}.
\end{equation}
We clearly see from the last expression that the energy is not bounded from
below for \( x_{0}\approx y_{0}\rightarrow \infty  \). The starting hypothesis
\( x_{0}\ll y_{0} \) is thus not satisfied. In the second case \( x_{0}\gg y_{0} \).
With the help of (\ref{eCoul>>res}) in the Appendix, the total energy per chain
is 
\begin{equation}
\label{EtotSup}
E_{tot}\approx -2w_{y}\log \left( \frac{x_{0}}{a_{x}}\right) -2w_{y}\frac{y_{0}}{x_{0}}\log \left( \frac{x_{0}}{y_{0}}\right) +4a_{y}\frac{J_{\perp }}{a_{x}}\frac{x_{0}}{y_{0}}.
\end{equation}
 Minimizing with respect to \( x_{0} \) and \( y_{0} \) leads to \( x_{0}\approx y_{0}\rightarrow \infty  \),
again in contradiction with the starting hypothesis \( x_{0}\gg y_{0} \). 

\noindent These contradictions imply that the modulated line minimizes its energy
at intermediate values of \( r_{0}=x_{0}/y_{0} \), e.g. \( r_{0}\approx 1 \),
where the Coulomb energy is of the order of the confinement energy. In this
regime, (\ref{TotalEnergy}) is difficult to treat analytically. 

\begin{figure}[tbph]
\begin{center}
\includegraphics[width=3.0in]{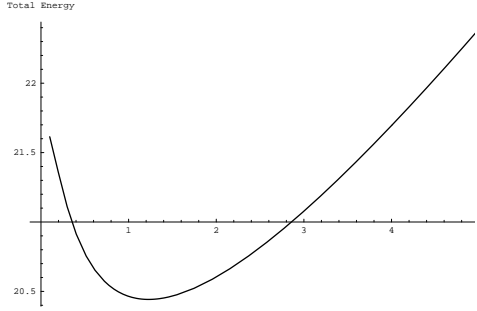}

\caption{\label{Fig.1}The total energy (\ref{TotalEnergy2}) with respect to \protect\( r_{0}\protect \). }
\end{center}
\end{figure}

\noindent We thus consider an ansatz giving the distribution of the kinks. Numerical
simulations give zig-zag shapes. In this connexion we show in the Appendix that
a sinusoidal profile 
\[
x(y)=x_{0}\sin (2\pi \frac{y}{y_{0}})\]
 gives a correct description of the distribution in the continuum limit. The
total energy is then given by
\begin{equation}
\label{TotalEnergy2}
E_{tot}=\frac{4}{\pi }w_{y}\int ^{l_{D}/a_{x}}_{a_{x}/y_{0}}dt\frac{1}{\sqrt{t^{2}+16\pi ^{2}r^{2}_{0}\sin ^{2}(t)}}K\left( \frac{4\pi r_{0}|\sin (t)|}{\sqrt{t^{2}+16\pi ^{2}r^{2}_{0}\sin ^{2}(t)}}\right) +4a_{y}\frac{J_{\perp }}{a_{x}}r_{0},
\end{equation}
where the upper cut-off corresponds to the dimensionless Debye length and \( K(k) \)
is the complete Elliptic function of the first kind. The lower cut-off in the
\( t- \)integral gives rise to the asymmetry between \( x_{0} \) and \( y_{0} \),
which is already present in (\ref{EtotInf}) and (\ref{EtotSup}). As the sinusoidal
distribution is valid in the continuum limit we shall take \( a_{x}\rightarrow 0 \)
and \( a_{y}\rightarrow 0 \) keeping other lengths fixed. (\ref{TotalEnergy2})
then depends only on \( r_{0} \) . The numerical plot of this energy clearly
shows a minimum for \( r_{0}\approx 1.2 \), cf. Fig. \ref{Fig.1}, which agrees
with the analytical predictions. 

With the help of (\ref{xopt/2DCoul}) these results lead to 
\begin{equation}
\label{param}
x_{0}\approx y_{0}\approx (w_{y}/J_{\perp })a_{x}\gg 1.
\end{equation}
The condition that the lines do not overlap, \( x_{0}\ll 1/\nu  \), returns
us to the boundary \( \Gamma <\Gamma _{c}=1/\nu  \) separating the phase of
linear aggregates from the Wigner crystal of bikinks.

\noindent A tilted stripe, which corresponds to a small argument of the sine
distribution above, is easily shown to better minimize the energy. This fact
is relevant in the context of cuprate oxides \cite{zaanen}.

\begin{figure}[tbph]
\begin{center}
\includegraphics[width=3.0in]{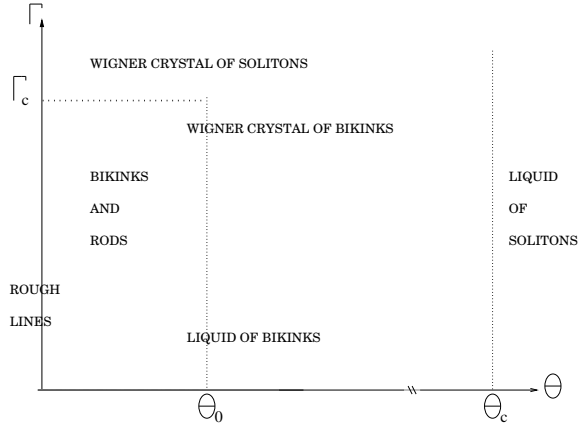}

\caption{\label{Fig2.phd}Phase diagram of the \protect\( 2D\protect \) model.}

\noindent \( \Theta _{c} \) is the confinement transition temperature below
which solitons aggregate into bikinks. \( \Theta _{0} \) is the crossover temperature
below which bikinks aggregate into growing rods. At \( \Theta =0 \) the latter
are roughened domain lines perpendicular to the chains when the Coulomb interaction
\( \Gamma <\Gamma _{c} \). For \( 0<\Theta <\Theta _{0} \) this confined phase
consists of aggregates screened by a liquid or a Wigner crystal of bikinks,
depending on the strength of the Coulomb interaction. When \( \Gamma >\Gamma _{c} \)
both the lines at \( \Theta =0 \) and the aggregates with their screening charges
at \( 0<\Theta <\Theta _{0} \) crossover to a deconfined phase, the Wigner
crystal of solitons.
\end{center}
\end{figure}

\section{Numerical approach}

\label{numerics}In this section we consider the results obtained with the help
of numerical methods. The calculations have been performed in the canonical
ensemble with periodic boundary conditions. The latter implies interaction of
each kink, within a computational cell, not only with all other kinks but also
with all their images residing in cells obtained by translation from the original
computational cell \cite{gron}. This results in an effective electrostatic
interaction which is periodic in the computational cell used and which can be
properly tabularized for rapid evaluations \cite{leckner}. At the beginning
of each simulation we place the kinks at random and assign the sign to all of
the Ising variables \( S_{n,\alpha } \) accordingly. We study the energy landscape
of the system using the classical Monte Carlo technique with the standard Metropolis
algorithm for the acceptance of kink motion. In order to rapidly reach configurations
with the lowest energy, we perform simulated annealing from high temperatures
with a variable sampling time range. We emphasize the common points between
these numerics and the various points discussed above. Also, the impact of discreteness
within the simulations is considered.

\subsection{\noindent The parameters of numerical simulations and the impact of the coarse
graining}

The typical values taken in numerical simulations are \( J_{\perp }\approx 10^{-1}eV \),
\( \nu \approx 1\% \) and \( a_{y}\approx 4,5A \). This leads to \( \Theta _{c}\approx 100 \),
\( \Theta _{0}\approx 1 \), as above, and \( \Gamma \approx 10/\epsilon  \).
The Coulomb interaction is then monitored with the help of the dielectric constant.
The value assigned to the coarse graining length needs some more care. It must
be at least of the order of the core of a soliton (\( \xi _{0}\approx 7A \)
in polyacetylene). However, if it is too large the ground state of the system
may be unreachable because of commensurability effects due to the discreteness
of the lattice. The latter are especially relevant in the Wigner crystal phases
where a pinning by the lattice occurs which distorts the Wigner crystal from
its equilibrium position. We shall therefore estimate the upper boundary, allowing
neglect of such commensurability effects, which can be reached by \( a_{x} \).

\noindent We suppose that kinks form a Wigner crystal with their positions \( x^{(eq)}_{n,\alpha } \)
in equilibrium, i.e. at the lowest energy distribution which is the triangular
lattice \cite{marad}. This structure melts when the fluctuations \( u \) of
the kinks around their equilibrium position is a fraction \( \gamma  \), the
Lindeman number, of their mean separation \( \sqrt{s/\nu } \). The new positions
are \( x_{n,\alpha }=x^{(eq)}_{n,\alpha }+u \), with \( u=\gamma \sqrt{s/\nu } \)
, \( \gamma <1 \). Usually the melting is due to thermal or quantum fluctuations.
However, if the coarse graining imposed by simulations is greater than \( u \),
kinks will also never reach their equilibrium position and the Wigner crystal
will not be observed. Instead a two-dimensional commensurate phase will form.
Thus, the coarse grained length must satisfy the condition \( a_{x}<u=\gamma \sqrt{\xi _{0}a_{y}/\nu } \).
With the typical values given above this yields \( a_{x}<50\gamma A \). For
this range of the coarse graining length the kinks will be able to adjust themselves
and not be influenced by the underlying lattice. Supposing that \( \gamma \approx 0,5 \)
gives an overestimated value of the maximum length, \( a_{x}\approx 25A \). 
\bigskip{}

\subsection{\noindent Numerical results}

\begin{figure}[tbph]
\begin{center}
\includegraphics{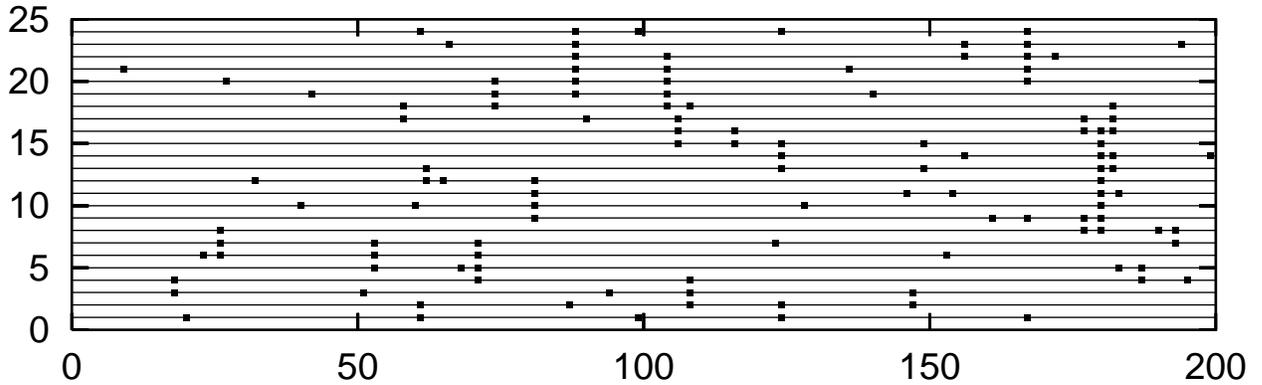}

\caption{\label{Fig.3rodp}The non-interacting rod phase corresponding to \protect\( 0\ll \Theta <\Theta _{0}\protect \). }

\noindent This phase has been obtained by imposing a large dielectric constant.
The computational cell displays the chains as horizontal lines, and their excitations
as the points. Each excitation corresponds to the presence of a reversed spin
with respect to the ferromagnetic ground state and thus represents a bikink.
The aggregation of bikinks leads to the presence of rods perpendicular to the
chains.
\end{center}
\end{figure}

\noindent Figure \ref{Fig.3rodp} shows the non-interacting phase at temperatures
\( 0\ll \Theta <\Theta _{0} \). This phase has been obtained from the interacting
case by imposing a large dielectric constant \( \epsilon \sim 10^{9} \) so
that \( \Gamma \ll 1 \). In accordance with what has been said concerning the
non-interacting phase, a gas of bikinks is present and the latter aggregates
into rods perpendicular to the chains.
 
\medskip{}

\begin{figure}[tbph]
\begin{center}
\includegraphics{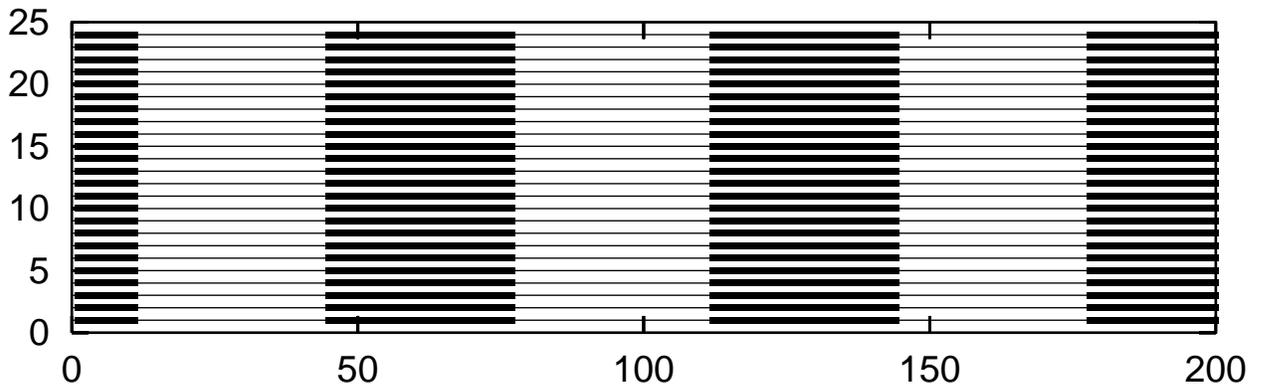}

\caption{\label{Fig.doml}Domain lines at \protect\( \Theta =0\protect \) for a weak
Coulomb interaction \protect\( \Gamma \approx 1\protect \).}

\noindent As temperature approaches zero the rods of figure \ref{Fig.3rodp}
cross the whole computational cell more like a system of domain lines. In the
presence of a weak Coulomb interaction these lines order in the form of a one
dimensional Wigner crystal.
\end{center}
\end{figure}

\noindent Turning on the Coulomb interactions and adjusting the dielectric constant
such that \( \Gamma \approx 1\ll \Gamma _{c} \), we have at \( \Theta =0 \)
a one dimensional Wigner crystal of domain lines as shown in figure \ref{Fig.doml}. 

\medskip{}

\begin{figure}[tbph]
\begin{center}
{\par\centering \includegraphics{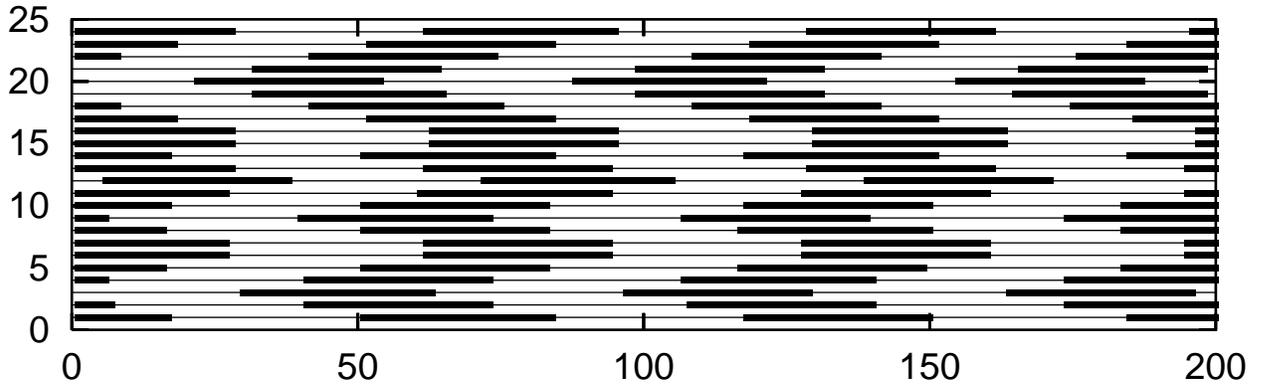} \par}

\caption{\label{zigzag}Modulated domain lines at \protect\( \Theta =0\protect \) and
\protect\( 1\ll \Gamma <\Gamma _{c}\protect \)}

\noindent A Coulomb roughening of the domain lines takes place when Coulomb
interaction is increased, in agreement with the results of paragraph \ref{weakCoul}.
\end{center}
\end{figure}

\noindent Turning on stronger Coulomb interactions the lines acquire a modulated
shape in accordance with what has been said in section \ref{weakCoul}; the
result is displayed in Fig. \ref{zigzag} for \( \Gamma \approx 10 \). By comparison
with Fig. \ref{Fig.doml} the lines are still forming a periodic pattern. They
roughen as a result of the competition between Coulomb interactions and confinement. 
 
\medskip{}

\section{Conclusion}

We have considered the effect of two competing long range interactions, Coulomb
and confinement, on the thermodynamic properties of systems with two-fold degenerate
ground-states. Analytical and numerical methods have been used and compared.
The theory has been developed along the lines of the non-interacting case where
a mapping to the Ising model allowed for an exact solution in \( 2D \). Due
to the complications brought by the Coulomb interactions, we have studied, in
a semi-phenomenological way, the main features which distinguish the present
case from the non-interacting one. In \( 2D \) the main parameter controlling
the aggregation of solitons in the presence of the long range interactions is
\( \Gamma  \) in equation (\ref{gamma}), i.e. the ratio between the Coulomb
energy scale and the confinement scale. This parameter, below the crossover
temperature \( \Theta _{0} \), cf. (\ref{teta0}), is analogous to temperature
above the confinement transition \( \Theta _{c} \), cf. (\ref{tetac}). The
various confined and deconfined phases have been displayed on the phase diagram
Fig. \ref{Fig2.phd} depending on \( \Gamma  \) and temperature. We have concentrated
on the low temperature regime , i.e. below \( \Theta _{0} \), where we have
shown that a crossover line \( \Gamma _{c}\sim 1/\nu  \) separates a confined
phase of bikinks and their aggregates from a deconfined phase which is a Wigner
crystal of solitons. As expected, in the aggregated phases, e.g. rods and lines,
the Coulomb interactions may play a crucial role because of the size dependent
charging energy of the solitonic complexes. In the case of weak Coulomb interactions
\( \Gamma \ll 1 \), the efficiency of the bulk screening insures the preservation
of the aggregates. The main effect of the long range interactions is to shift
the chemical potential of the elementary defects by the Coulomb self-energy
(\ref{Kpara*}). However, when bulk screening is no longer efficient, we have
found that antiferromagnetic interactions along the chains are necessary in
order for the system to adjust the correct density of solitons and conserve
the aggregates (\ref{nnonscreen}). The ground state of the system remains ferromagnetic
to maintain the electroneutrality. In both cases, efficient and weak screening,
a growing macroscopic aggregation takes place. Because of thermal roughening,
only at \( T=0 \) do the complexes of solitons span the whole sample in the
form of domain lines (\ref{SOSCorrF}). We have shown that the latter are stable
for larger Coulomb interactions \( 1\ll \Gamma <\Gamma _{c} \) due to self-screening
in the ensemble of solitons. This is manifested by the Coulomb roughening of
the lines in accordance with numerical results, cf. Fig. \ref{zigzag}. The
parameters of the distribution were derived from a variational procedure (\ref{param}).
In the continuum limit a sinusoidal profile gives a correct description of the
rough lines. 

\section*{Appendix }

\noindent \label{distrofsolitons} We consider the periodic modulation for a
general distribution of kinks \( x(y) \). The Coulomb potential energy associated
with this distribution is given by (\ref{poten}). Going to the new variables
\( t=(y-y')/2 \) and \( s=(y+y')/2 \) expression (\ref{poten}) becomes
\begin{equation}
\label{eCoulEnergy}
E_{coul}=\frac{2w_{y}}{L}\int ^{L}_{a_{x}}ds\int ^{L}_{a_{x}}\frac{dt}{a_{y}}\left\{ \frac{1}{\sqrt{t^{2}+\left( x(s+t)-x(s-t)\right) ^{2}}}-\frac{1}{t}\right\} .
\end{equation}
 The integral has been regularized by subtracting the infinite contribution
coming from the straight line \( x(y)\equiv 0 \). 

\bigskip{}
\noindent In the case where \( x_{0}\ll y_{0} \), (\ref{eCoulEnergy}) can
be divided as follows:
\begin{equation}
\label{eCoul<<}
E_{coul}=\frac{2w_{y}}{L}\int ^{L}_{a_{x}}ds\left[ \int ^{x_{0}}_{a_{x}}+\int ^{y_{0}}_{x_{0}}+\int ^{L}_{y_{0}}\right] \frac{dt}{a_{y}}\left\{ \frac{1}{\sqrt{t^{2}+\left( x(s+t)-x(s-t)\right) ^{2}}}-\frac{1}{t}\right\} .
\end{equation}
In all integrals in (\ref{eCoul<<}) we can expand \( x(s\pm t)\approx x(s)\pm tx'(s) \).
Thus the \( t \) and \( s \) integrals decouple and we obtain
\begin{equation}
\label{eCoul<<res}
E_{coul}\approx -2w_{y}\left( \frac{x_{0}}{y_{0}}\right) ^{2}\log \left( \frac{y_{0}}{a_{x}}\right) .
\end{equation}
In the opposite case, \( x_{0}\gg y_{0} \), we have
\begin{equation}
\label{eCoul>>}
E_{coul}\approx \frac{2w_{y}}{L}\int ^{L}_{a_{x}}ds\left[ \int _{a_{x}}^{x_{0}}+\int ^{L}_{x_{0}}\right] \frac{dt}{a_{y}}\left\{ \frac{1}{\sqrt{t^{2}+\left( x(s+t)-x(s-t)\right) ^{2}}}-\frac{1}{t}\right\} .
\end{equation}
In the present case the expansion of \( x(s\pm t) \) needs some care because
\( x_{0}\gg y_{0} \) and the \( s \) and \( t \) integral do not easily decouple.
The first \( t- \)integral, in (\ref{eCoul>>}), is over \( x_{0}/y_{0}\gg 1 \)
periods. Its major contribution is at the vicinity of the zeros of \( x(s+t)-x(s-t) \).
These zeros exist for any \( s \) and correspond to ``nesting'' points \( t_{n}\equiv ny_{0}/2 \)
with \( n=1,2,...2r_{0} \). From these considerations and with the help of
\( \delta t=t-t_{n} \), the major contribution to the integral reads
\[
2w_{y}\sum ^{2r_{0}}_{n=1}\int ^{\frac{y_{0}}{2}}_{0}\frac{d\delta t}{a_{y}}\frac{1}{\sqrt{t_{n}^{2}+4\delta t^{2}x'^{2}(s)}}\approx \frac{2w_{y}}{|x'(s)|}\sum ^{2r_{0}}_{n=1}\log \left( \frac{4x'(s)}{n}\right) \approx -2w_{y}\int ^{1}_{\frac{1}{r_{0}}}dn\log (n),\]
where \( r_{0}=x_{0}/y_{0} \) and we have taken \( x'(s)\approx r_{0} \).
Thus, restricting ourselves to the main logarithmic approximation, we obtain
\[
-2w_{y}\log \left( \frac{x_{0}}{a_{x}}\right) -2w_{y}\frac{y_{0}}{x_{0}}\log \left( \frac{x_{0}}{y_{0}}\right) .\]
In the second \( t- \)integral \( t \) is large and we can expand the denominator
in \( x(s+t)-x(s-t) \). Going to the first order in \( 1/r_{0} \) the result
is compensated by the regularizing term and thus gives zero.

\noindent The Coulomb energy then reads 
\begin{equation}
\label{eCoul>>res}
E_{coul}\approx -2w_{y}\frac{y_{0}}{x_{0}}\log \left( \frac{x_{0}}{y_{0}}\right) -2w_{y}\log \left( \frac{x_{0}}{a_{x}}\right) .
\end{equation}

\noindent The total energy of the system we are considering includes the potential
energy and the confinement energy. The latter can be taken, for any \( x_{0} \),
as
\begin{equation}
\label{econf}
E_{conf}=a_{y}\frac{J_{\perp }}{a_{x}}\int ^{y_{0}}_{0}|\frac{dx}{dy}|dy=4a_{y}\frac{J_{\perp }}{a_{x}}\frac{x_{0}}{y_{0}}
\end{equation}
Combining (\ref{econf}) with either (\ref{eCoul<<res}) or (\ref{eCoul>>res})
the total energy per chain is thus given by (\ref{TotalEnergy}).

\begin{figure}[tbph]
\begin{center}
\includegraphics[width=3.0in]{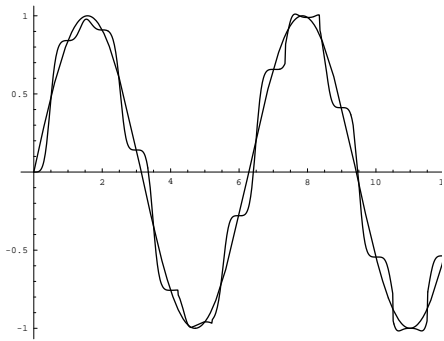}

\caption{\label{Fig3}The sinusoidal distribution.}
\end{center}
\end{figure}

\vfill{}
\noindent A particular function \( x(y) \) describing a single domain line
in the two-dimensional plane at \( T=0 \) can be estimated. As implicitly assumed
in (\ref{econf}), the confinement of this aggregate can be taken into account,
in the Hamiltonian, with the help of a solid on solid model: 
\begin{equation}
\label{Hinitial}
H=\frac{1}{2\epsilon }\sum _{y,y'}\frac{e^{2}}{\sqrt{(x(y)-x(y'))^{2}+(y-y')^{2}}}+V\sum _{y}|x(y)-x(y-a_{y})|.
\end{equation}
 Minimizing (\ref{Hinitial}) with respect to \( x(y) \) yields the recurrence
relation
\begin{equation}
\label{recurrencex(y)}
x(y)=\frac{\kappa ^{2}\left( sign\left( x(y)-x(y-a_{y})\right) -sign\left( x(y+a_{y})-x(y)\right) \right) +\sum _{y'}x(y')\left[ \left( x(y)-x(y')\right) ^{2}+\left( y-y'\right) ^{2}\right] ^{-3/2}}{\sum _{y'}\left[ \left( x(y)-x(y')\right) ^{2}+\left( y-y'\right) ^{2}\right] ^{-3/2}},
\end{equation}
where \( sign \) is the sign function and \( \kappa ^{2}=\epsilon V/e^{2}=(\xi _{0}a_{y}\Gamma )^{-1} \),
has a unit of inverse length squared and is related to (\ref{xopt/2DCoul}).
The solution \( x(y) \) cannot easily be found analytically. However simulations
give us a hint for they show a kind of zig-zag distribution. This suggests using
a sinusoidal ansatz
\begin{equation}
\label{x(y)}
x(y)=x_{0}\sin (2\pi \frac{y}{y_{0}}).
\end{equation}
Comparing the resulting plot of each side of (\ref{recurrencex(y)}) we can
check the accuracy of this approximation. The result is displayed in Figure
(\ref{Fig3}), which shows that the sinusoidal distribution is correct in the
continuum limit. 

\end{document}